\documentclass[12pt]{article}

\usepackage{graphicx}

\def\ltwid{\mathrel{\raise.3ex\hbox{$<$\kern-.75em\lower1ex\hbox{$\sim$}}}}
\def\gtwid{\mathrel{\raise.3ex\hbox{$>$\kern-.75em\lower1ex\hbox{$\sim$}}}}

\def\square{\kern1pt\vbox{\hrule height 1.2pt\hbox{\vrule width 1.2pt\hskip 3pt
   \vbox{\vskip 6pt}\hskip 3pt\vrule width 0.6pt}\hrule height 0.6pt}\kern1pt}
\def\overleftrightarrow#1{\vbox{\ialign{##\crcr
     $\leftrightarrow$\crcr\noalign{\kern-1pt\nointerlineskip}
     $\hfil\displaystyle{#1}\hfil$\crcr}}}

\begin{document}

\begin{titlepage}

\begin{flushright}
CCTP-2012-08 \\ UFIFT-QG-12-04
\end{flushright}

\vskip 2cm

\begin{center}
{\bf Weyl-Weyl Correlator in de Donder Gauge on de Sitter}
\end{center}

\vskip 1cm

\begin{center}
P. J. Mora$^{1*}$, N. C. Tsamis$^{2\dagger}$ and R. P.
Woodard$^{1,\ddagger}$
\end{center}

\vskip 1cm

\begin{center}
\it{$^{1}$ Department of Physics, University of Florida \\
Gainesville, FL 32611, UNITED STATES}
\end{center}

\begin{center}
\it{$^{2}$ Institute of Theoretical Physics \& Computational Physics \\
Department of Physics University of Crete \\
GR-710 03 Heraklion, HELLAS}
\end{center}

\vskip 1cm

\begin{center}
ABSTRACT
\end{center}
We compute the linearized Weyl-Weyl correlator using a new
solution for the graviton propagator on de Sitter background
in de Donder gauge. The result agrees exactly with a previous
computation in a noncovariant gauge. We also use dimensional
regularization to compute the one loop expectation value of
the square of the Weyl tensor.

\begin{flushleft}
\end{flushleft}

\vskip 1cm

\begin{flushleft}
$^*$ e-mail: pmora@phys.ufl.edu \\
$^{\dagger}$ e-mail: tsamis@physics.uoc.gr \\
$^{\ddagger}$ e-mail: woodard@phys.ufl.edu
\end{flushleft}

\end{titlepage}

\section{Introduction}\label{intro}

De Sitter space holds great phenomenological interest as a paradigm
for the background geometry of primordial inflation. From this
perspective there is no point to working on the full de Sitter
manifold or paying any special account to the full de Sitter group.
The cosmological patch of de Sitter is merely the special case of a
spatially flat, Friedman-Robertson-Walker geometry whose Hubble
parameter happens to be exactly constant, and the important
symmetries are homogeneity and isotropy. In stark contrast,
mathematical physicists accord the de Sitter geometry a special
status as the unique maximally symmetric solution to the Einstein
equations with a positive cosmological constant. They believe
strongly that the full de Sitter group should play the same role in
organizing quantum field theory on de Sitter as the Poincar\'e group
does for flat space.

It should be emphasized that quantum field theories do what they
please, without regard for the prejudices of those who study them.
So de Sitter invariance is recovered, when it is present, whether or
not explicit account is taken of it. This is exactly what happens if
one constructs the Bunch-Davies mode sums (in $D$ spacetime
dimensions with Hubble parameter $H$) for the propagators of a
minimally coupled scalar with $M_S^2 > 0$ \cite{CT}, a spin one half
fermion with $M_F^2 > -\frac{D}2 H^2$ \cite{CR}, or a transverse
vector with $M_V^2 > -2(D-1) H^2$ \cite{MTW1}. However, infrared
divergences break de Sitter invariance if the mass-squared drops
below these bounds \cite{MTW1,AF}.\footnote{The reason for this is
obvious: each mode is an independent harmonic oscillator, and making
the mass-squared drop below the stated bounds results in a potential
which curves downwards. So the mode tends to roll down its potential.
If the quantum state is released centered about the origin, it will
spread, and how far it spreads depends upon when it was released.
Mathematical physicists sometimes deny this by resorting to analytic
regularization schemes which automatically subtract off power law
infrared divergences. This results in the curious claim that
tachyonic scalars have de Sitter invariant propagators, except for
the discrete values $M^2 = -N (N + D-1) H^2$ \cite{Higuchi}. The
special thing about these masses is that they make a formerly power
law infrared divergence logarithmic, and hence visible to the analytic
regularization \cite{MTW1}. In fact, it is never correct to subtract
off infrared divergences, and the subtracted mode sums which result
from this bogus procedure solve the propagator equation without
being true propagators \cite{TW1}.}

The case of gravitons has long been recognized as dynamically
equivalent to that of massless, minimally coupled scalars
\cite{Grishchuk}. Hence there can be no de Sitter invariant graviton
propagator. For this reason a noninvariant gauge fixing functional
was employed to construct the only graviton propagator
\cite{TW2,RPW} for which any loop computations have been performed
\cite{TW3,TW4,TW5,MW1,KW1}. Although this propagator breaks de
Sitter invariance because of its gauge fixing functional, the method
of the compensating gauge transformation reveals a physical
violation of de Sitter invariance as well \cite{Kleppe}.

Mathematical physicists disputed this conclusion for many years
because adding de Sitter invariant gauge fixing terms to the action
results in a propagator equation with de Sitter invariant solutions
\cite{INVPROP}. However, it has recently been shown that there is a
topological obstacle to adding invariant gauge fixing functionals on
any manifold, such as de Sitter, which possesses a linearization
instability \cite{MTW2}. Ignoring this problem for scalar quantum
electrodynamics on de Sitter leads to on shell singularities in the
one loop scalar self-mass-squared \cite{KW2}, and would cause similar
problems were it done in quantum gravity. It is still possible to
impose invariant gauges which are ``exact'' in the sense that the
field obeys some strong operator equation. One would naively think
that exact gauge conditions could be obtained by taking singular
limits of gauge fixing terms \cite{Higuchi}, but this step involves
analytic continuation in a gauge parameter, which is highly suspect
when infrared divergences are present \cite{MTW1}. The more reliable
technique is simply to construct the propagator directly in the exact
gauge. This has been done recently for de Donder gauge \cite{MTW3},
and the resulting propagator shows de Sitter breaking in both the
spin two and spin zero sectors \cite{MTW3,KMW}.

Continued resistance to the reality of de Sitter breaking in quantum
gravity has been based on three arguments:
\begin{itemize}
\item{That de Sitter invariant propagators result from taking
certain limits of gauge fixing terms which naively enforce
exact gauge conditions \cite{FH};}
\item{That the infrared divergence which precludes a de Sitter
invariant propagator for dynamical gravitons is a gauge artifact
\cite{HMM}; and}
\item{That previous de Sitter breaking solutions for the graviton
propagator \cite{TW2,RPW,MTW3,KMW} are in a different sector of
the Hilbert space \cite{HMM}.}
\end{itemize}
We have already explained that the first argument involves a dubious
analytic continuation. (It would be interesting to check if acting
the graviton kinetic operator on the claimed propagators produces
the appropriate functional projection operators.) The second
argument has been rebutted by demonstrating that the putative gauge
transformation amounts to an alternate quantization scheme which
changes physical quantities such as the tensor power spectrum
\cite{MTW4}. The final argument was partially answered by computing
the linearized Weyl-Weyl correlator for the first of the two de Sitter
breaking propagators \cite{MW}. The result for this that had previously
been accepted by mathematical physicists \cite{Kouris} turns out to
contain some mistakes \cite{MW}, but correcting these leads to complete
agreement with the de Sitter breaking propagator \cite{Atsuchi}. We do
not believe that the Weyl-Weyl correlator completely checks the
graviton propagator, but mathematical physicists ascribe great
significance to it \cite{HMM}, and the result certainly undermines
their skepticism about the first of the two de Sitter breaking
propagators.

The purpose of this paper is to finish answering the third argument
by computing the linearized Weyl-Weyl correlator for de Sitter
breaking propagator recently derived in exact de Donder gauge
\cite{MTW3,KMW}. Section \ref{previous} reviews a number of
technical results we shall need from previous work. The actual
computation is performed in section \ref{compute}. Our conclusions
comprise section \ref{discuss}.

\section{Notation and Previous Work}\label{previous}

The purpose of this section is to explain notation and introduce
certain key results from previous work which facilitate the present
study. We begin with results from the recent computation of the
Weyl-Weyl correlator for the older of the two de Sitter breaking
propagators \cite{MW}. Then the de Donder gauge propagator is
described \cite{MTW3,KMW}.

\subsection{For the Weyl-Weyl Correlator}\label{weyl}

We work on the ``cosmological patch'' of $D$-dimensional de Sitter,
which can be covered using conformal coordinates $x^{\mu} =
(\eta,\vec{x})$ with,
\begin{equation}
-\infty < \eta < 0 \qquad , \qquad -\infty < x^i < +\infty \qquad
{\rm for} \qquad i = 1, \ldots, D\!-\!1 \; .
\end{equation}
The metric in these coordinates is conformal to that of flat space,
\begin{equation}
ds^2 = a^2 \Bigl( -d\eta^2 + d\vec{x} \!\cdot\! d\vec{x} \Bigr)
\qquad {\rm where} \qquad a \equiv -\frac1{H \eta} \; .
\end{equation}
The parameter $H$ is known as the Hubble constant, and is related
to the cosmological constant by $\Lambda = (D-1) H^2$.

It is convenient to represent the propagator between points $x^{\mu}$
and ${x'}^{\mu}$ using the de Sitter length function $y(x;x')$,
\begin{equation}
y(x;x') \equiv a a' H^2 \Biggl[\Bigl\Vert \vec{x} \!-\! \vec{x}'
\Vert^2 - \Bigl( \vert \eta \!-\! \eta' \vert \!-\! i \varepsilon
\Bigr)^2 \Biggr] \; . \label{ydef}
\end{equation}
Except for the factor of $i \varepsilon$ (whose purpose is to
enforce Feynman boundary conditions) the de Sitter length function
can be expressed as follow in terms of the geodesic length
$\ell(x;x')$ from $x^{\mu}$ to ${x'}^{\mu}$,
\begin{equation}
y(x;x') = 4 \sin^2\Bigl( \frac12 H \ell(x;x')\Bigr) \; .
\end{equation}
When de Sitter invariance cannot be maintained, we elect to preserve
homogeneity and isotropy --- this is known as the ``E(3)'' vacuum
\cite{BA}. This means the propagator depends upon $y(x;x')$ and the
scale factors at $x^{\mu}$ and ${x'}^{\mu}$. An important example is
the propagator of the massless, minimally coupled scalar \cite{OW},
\begin{equation}
i\Delta_A(x;x') = A\Bigl( y(x;x') \Bigr) + k \ln(a a') \; , \label{DA}
\end{equation}
where the constant $k$ is,
\begin{equation}
k \equiv \frac{ H^{D-2}}{(4 \pi)^{\frac{D}2}} \frac{\Gamma(D \!-\! 1)}{
\Gamma(\frac{D}2)} \; .
\end{equation}
The function $A(y)$ is,
\begin{eqnarray}\label{Af}
\lefteqn{A(y) = \frac{H^{D-2}}{(4\pi)^{\frac{D}2}} \Biggl\{
\Gamma\Bigl(\frac{D}2 \!-\!1\Bigr) \Bigl(\frac{4}{y}\Bigr)^{
\frac{D}2 -1} \!+\! \frac{\Gamma(\frac{D}2 \!+\! 1)}{\frac{D}2
\!-\! 2} \Bigl(\frac{4}{y} \Bigr)^{\frac{D}2-2} \!+\! A_1 }
\nonumber \\
& & \hspace{1.5cm} - \sum_{n=1}^{\infty} \Biggl[
\frac{\Gamma(n\!+\!\frac{D}2\!+\!1)}{(n\!-\!\frac{D}2\!+\!2) (n
\!+\! 1)!} \Bigl(\frac{y}4 \Bigr)^{n - \frac{D}2 +2} \!\!\!\!\! -
\frac{\Gamma(n \!+\! D \!-\! 1)}{n \Gamma(n \!+\! \frac{D}2)}
\Bigl(\frac{y}4 \Bigr)^n \Biggr] \Biggr\} , \qquad \label{DeltaA}
\end{eqnarray}
where the constant $A_1$ is,
\begin{equation}
A_1 = \frac{\Gamma(D\!-\!1)}{\Gamma(\frac{D}2)} \Biggl\{
-\psi\Bigl(1 \!-\! \frac{D}2\Bigr) + \psi\Bigl(\frac{D\!-\!1}2\Bigr)
+
\psi(D \!-\!1) + \psi(1) \Biggr\} .
\end{equation}

Two permutations have great importance for us: {\it Riemannization}
and {\it Weylization}. The first was originally introduced as ``the
standard permutation'' in a study of one loop corrections to an
invariant correlator of two Riemann tensors on flat space background
\cite{TW6}. It operates on any 8-index bi-tensor ``seed''
$S_{\alpha\beta\gamma\delta\mu\nu\rho\sigma}$ with the algebraic
symmetries of a graviton propagator with two ordinary derivatives at
each point:
\begin{equation}
S_{\alpha\beta\gamma\delta\mu\nu\rho\sigma} =
S_{\beta\alpha\gamma\delta\mu\nu\rho\sigma} =
S_{\alpha\beta\delta\gamma\mu\nu\rho\sigma} =
S_{\alpha\beta\gamma\delta\nu\mu\rho\sigma} =
S_{\alpha\beta\gamma\delta\mu\nu\sigma\rho} \; .
\end{equation}
Riemannization permutes the seed tensor so that it has the algebraic
symmetries of the product of two linearized Riemann tensors,
\begin{equation}
\label{Rdef} {\rm Riem}\Bigl[ S_{\alpha\beta\gamma\delta
\mu\nu\rho\sigma} \Bigr] \equiv
\mathcal{R}_{\alpha\beta\gamma\delta}^{~~~~~
\epsilon\zeta\kappa\lambda} \times
\mathcal{R}_{\mu\nu\rho\sigma}^{~~~~~ \theta\phi\psi\omega} \times
S_{\epsilon\zeta\kappa\lambda\theta\phi\psi\omega} \; ,
\end{equation}
 where,
\begin{equation}
\mathcal{R}_{\alpha\beta\gamma\delta}^{~~~~~\epsilon\zeta\kappa\lambda}
\equiv \delta^{\epsilon}_{\alpha} \delta^{\kappa}_{\gamma}
\delta^{\zeta}_{\beta} \delta^{\lambda}_{\delta}
- \delta^{\epsilon}_{\gamma} \delta^{\kappa}_{\beta}
\delta^{\zeta}_{\delta} \delta^{\lambda}_{\alpha}
+ \delta^{\epsilon}_{\beta} \delta^{\kappa}_{\delta}
\delta^{\zeta}_{\alpha} \delta^{\lambda}_{\gamma}
- \delta^{\epsilon}_{\delta} \delta^{\kappa}_{\alpha}
\delta^{\zeta}_{\gamma} \delta^{\lambda}_{\beta} \; . \label{scriptR}
\end{equation}
Weylization operates on any bi-tensor seed with the algebraic
symmetries of the product of two Riemann tensors,
\begin{eqnarray}
S_{\alpha\beta\gamma\delta\mu\nu\rho\sigma}(x;x') & = &
S_{\gamma\delta\alpha\beta\mu\nu\rho\sigma}(x;x') =
S_{\mu\nu\rho\sigma\alpha\beta\gamma\delta}(x';x) \; , \\
S_{(\alpha\beta)\gamma\delta\mu\nu\rho\sigma}(x;x') & = & 0 =
S_{\alpha [\beta\gamma\delta]\mu\nu\rho\sigma}(x;x') \; .
\end{eqnarray}
Weylization subtracts off the traces within each index group to
produce something with the algebraic symmetries of the product of
two Weyl tensors,
\begin{equation}
{\rm Weyl}\Bigl[ S_{\alpha\beta\gamma\delta \mu\nu\rho\sigma}(x;x')
\Bigr] \equiv \mathcal{C}_{\alpha\beta\gamma\delta}^{~~~~~
\epsilon\zeta\kappa\lambda}(x) \times
\mathcal{C}_{\mu\nu\rho\sigma}^{~~~~~ \theta\phi\psi\omega}(x')
\times S_{\epsilon\zeta\kappa\lambda\theta\phi\psi\omega}(x;x') \; .
\label{Weylization}
\end{equation}
where,
\begin{eqnarray}
\label{cDef} \lefteqn{ \mathcal{C}_{\alpha\beta\gamma\delta}^{~~~~~
\epsilon \zeta\kappa\lambda} \equiv \delta^{\epsilon}_{\alpha}
\delta^{\zeta}_{\beta} \delta^{\kappa}_{\gamma}
\delta^{\lambda}_{\delta} - \Bigl[g_{\alpha\gamma}
\delta^{\zeta}_{\beta} \delta^{\lambda}_{\delta} - g_{\gamma\beta}
\delta^{\zeta}_{\delta} \delta^{\lambda}_{\alpha} + g_{\beta\delta}
\delta^{\zeta}_{\alpha} \delta^{\lambda}_{\gamma} - g_{\delta\alpha}
\delta^{\zeta}_{\gamma} \delta^{\lambda}_{\beta}
\Bigr] \frac{ g^{\epsilon\kappa}}{D \!-\! 2} } \nonumber \\
& & \hspace{6cm} + \Bigl[ g_{\alpha\gamma} g_{\beta\delta} \!-\!
g_{\alpha\gamma} g_{\beta\delta}\Bigr] \frac{g^{\epsilon\kappa}
g^{\zeta\lambda}}{(D \!-\! 2) (D \!-\!1) } \; . \qquad \label{scriptC}
\end{eqnarray}

The older of the two de Sitter breaking propagators \cite{TW2,RPW}
leads to the following result for the Weyl-Weyl correlator
\cite{MW},
\begin{eqnarray}
\lefteqn{ \Bigl\langle \Omega \Bigl\vert
C_{\alpha\beta\gamma\delta}(x) \times C_{\mu\nu\rho\sigma}(x')
\Bigr\vert \Omega \Bigr\rangle = 4 \pi G \, {\rm Weyl}\Biggl( {\rm
Riem}\Biggl[ D_{\alpha} D_{\gamma} D_{\mu}' D_{\rho}' \, i
\Delta_A(x;x') } \nonumber \\
& & \hspace{1cm} \times \Bigl[ \mathcal{R}_{\beta\nu}(x;x')
\mathcal{R}_{\delta\sigma}(x;x') \!+\!
\mathcal{R}_{\beta\sigma}(x;x') \mathcal{R}_{\delta\nu}(x;x') \Bigr]
\Biggr] \Biggr) + O(G^2) \; . \qquad \label{oldweyl}
\end{eqnarray}
Here and henceforth, $G$ is Newton's constant, $D_{\mu}$ stands for
the covariant derivative (in the de Sitter background) with respect
to $x^{\mu}$, $D'_{\rho}$ denotes the covariant derivative with
respect to ${x'}^{\rho}$, and $\mathcal{R}_{\mu\nu}(x;x')$ is the de
Sitter invariant, mixed partial derivative of $y(x;x')$, normalized
so that its flat space limit gives $\eta_{\mu\nu}$,
\begin{equation}
\mathcal{R}_{\mu\nu}(x;x') \equiv -\frac1{2 H^2} \frac{\partial^2
y(x;x')}{\partial x^{\mu} \partial {x'}^{\nu}} \; .
\end{equation}
Because the computation was done in $D$ spacetime dimensions, it is
a simple matter to take the coincidence limit using dimensional
regularization and contract the indices together \cite{MW},
\begin{equation}
\label{Csquared} \Bigl\langle \Omega \Bigl\vert
C^{\alpha\beta\gamma\delta}(x) C_{\alpha\beta\gamma\delta}(x)
\Bigr\vert \Omega \Bigr\rangle \!=\! 64\pi (D\!-\!3) D (D \!+\! 1)
(D\!+\! 2) A''(0) G H^4 \!+\! O(G^2 H^8) .
\end{equation}
The coincidence limit of the second derivative of $A(y)$ is,
\begin{equation}
\label{climit2} A''(0) = \frac{H^{D-2}}{(4\pi)^{\frac{D}2}} \times
\frac1{16} \frac{\Gamma(D \!+\! 1)}{\Gamma(\frac{D}2 \!+\! 2) } \; .
\end{equation}

\subsection{For the de Donder Gauge Propagator}

The graviton propagator in de Donder gauge can be expressed as the
sum of a spin zero part and a spin two part,
\begin{equation}
i \Bigl[\mbox{}_{\alpha\beta} \Delta_{\rho\sigma} \Bigr](x;x') = i
\Bigl[\mbox{}_{\alpha\beta} \Delta^0_{\rho\sigma} \Bigr](x;x') + i
\Bigl[\mbox{}_{\alpha\beta} \Delta^2_{\rho\sigma} \Bigr](x;x') \; .
\label{gravdecomp}
\end{equation}
Each part is represented as the product of differential projectors
that enforce the de Donder condition on each coordinate, acting
on a scalar structure function. For the spin zero part this form is,
\begin{equation}
i\Bigl[\mbox{}_{\mu\nu} \Delta^0_{\rho\sigma}\Bigr](x;x') =
\mathcal{P}_{\mu\nu}(x) \times \mathcal{P}_{\rho\sigma}(x')
\Bigl[\mathcal{S}_0(x;x') \Bigr] \; . \label{Spin0}
\end{equation}
The spin zero projector $\mathcal{P}_{\mu\nu}$ is a sum of longitudinal
and trace terms,
\begin{equation}
\mathcal{P}_{\mu\nu} \equiv D_{\mu} D_{\nu} + \frac{g_{\mu\nu}}{D
\!-\!2} \Bigl[ \square \!+\! 2 (D \!-\! 1) H^2 \Bigr] \; .
\label{spin0op}
\end{equation}
The spin two part takes the form,
\begin{equation}
i\Bigl[\mbox{}_{\mu\nu} \Delta^2_{\rho\sigma}\Bigr](x;z) = \frac1{4
H^4} \mathbf{P}_{\mu\nu}^{~~\alpha\beta}(x) \times
\mathbf{P}_{\rho\sigma}^{~~\kappa\lambda}(x') \Bigl[
\mathcal{R}_{\alpha\kappa}(x;x') \mathcal{R}_{\beta\lambda}(x;x')
\mathcal{S}_2(x;x')\Bigr] \; . \label{Spin2}
\end{equation}
The spin two projector $\mathbf{P}_{\mu\nu}^{~~\alpha\beta}$ is,
\begin{eqnarray}
\lefteqn{\mathbf{P}_{\mu\nu}^{~~\alpha\beta} = \frac12
\Bigl(\frac{D\!-\!3}{D\!-\!2}\Bigr) \Biggl\{ -\delta^{\alpha}_{(\mu}
\delta^{\beta}_{\nu)} \Bigl[\square \!-\! D H^2\Bigr] \Bigl[ \square
\!-\! 2 H^2\Bigr] + 2 D_{(\mu} \Bigl[ \square \!+\! H^2\Bigr]
\delta^{(\alpha}_{\nu)} D^{\beta)} } \nonumber \\
& & \hspace{.3cm} - \Bigl( \frac{D \!-\!2}{D \!-\!1} \Bigr) D_{(\mu}
D_{\nu)} D^{(\alpha} D^{\beta)} + g_{\mu\nu} g^{\alpha\beta} \Bigl[
\frac{\square^2}{D \!-\! 1} \!-\! H^2 \square \!+\! 2 H^4\Bigr]
\qquad \nonumber \\
& & \hspace{.3cm} -\frac{D_{(\mu} D_{\nu)} }{D \!-\! 1} \Bigl[
\square \!+\! 2 (D \!-\! 1) H^2\Bigr] g^{\alpha\beta}
-\frac{g_{\mu\nu} }{D \!-\! 1} \Bigl[ \square \!+\! 2 (D \!-\! 1)
H^2\Bigr] D^{(\alpha} D^{\beta)} \Biggr\} . \qquad \label{spin2op}
\end{eqnarray}
It is transverse and traceless on each index group,
\begin{eqnarray}
g^{\mu\nu} \mathbf{P}_{\mu\nu}^{~~\alpha\beta} = & 0 & =
\mathbf{P}_{\mu\nu}^{~~\alpha\beta} g_{\alpha\beta} \; , \\
D^{\mu} \mathbf{P}_{\mu\nu}^{~~\alpha\beta} = & 0 & =
\mathbf{P}_{\mu\nu}^{~~\alpha\beta} D_{\alpha} \; .
\end{eqnarray}

A key identity concerns the result of acting either
$\mathbf{P}_{\mu\nu}^{~~\alpha\beta}(x)$ or $\mathbf{P}_{\rho\sigma}^{
\kappa\lambda}(x')$ on $\mathcal{R}{\alpha\kappa}(x;x')
\mathcal{R}_{\beta\lambda}(x;x')$ times a de Sitter invariant
structure function. If the operator is acted by itself then the
result is complicated, as expression (\ref{spin2op}) indicates.
However, {\it if the longitudinal and trace parts are projected out on
the other index group,} then a very simple result pertains \cite{KMW},
\begin{eqnarray}
\lefteqn{ \mathbf{P}_{\mu\nu}^{~~\alpha\beta}(x) \times
\mathbf{P}_{\rho\sigma}^{~~\kappa\lambda}(x') \Bigl[
\mathcal{R}_{\alpha \kappa} \mathcal{R}_{\beta\lambda} F(y) \Bigr] }
\nonumber \\
& & \hspace{1.9cm} = -\frac12 \Bigl( \frac{D \!-\! 3}{D \!-\! 2}\Bigr)
\mathbf{P}_{\rho\sigma}^{~~\kappa\lambda}(x') \Biggl[
\mathcal{R}_{\mu \kappa} \mathcal{R}_{\nu \lambda} \square \Bigl[
\square \!-\! (D \!-\! 2) H^2\Bigr] F(y) \Biggr] \; . \qquad
\label{1stPsimp}
\end{eqnarray}
In deriving an explicit form for the propagator, it was not possible
to use this identity a second time, to simplify the action of
$\mathbf{P}_{\rho\sigma}^{~~\kappa\lambda}(x')$, because no
operator remains to project out longitudinal and trace parts on the
index group at $x^{\mu}$. However, we will see that Riemannization
and Weylization provide the crucial projections, which allows the
identity to be used twice in computing the Weyl-Weyl correlator.

The identity (\ref{1stPsimp}) is so crucial because the spin two
structure function obeys the relation,
\begin{equation}
\square^2 \Bigl[ \square \!-\! (D \!-\! 2) H^2\Bigr]^2
\mathcal{S}_2(x;x') = 32 H^4 \Bigl(\frac{D \!-\! 2}{D \!-\!
3}\Bigr)^2 i\Delta_A(x;x') \; . \label{S2simp}
\end{equation}
While the spin two structure function is, by itself, very complicated
\cite{MTW3,KMW}, the action of precisely the derivatives in
(\ref{S2simp}) reduces it to the propagator of a massless, minimally
coupled scalar. We shall not require the spin zero structure function
but its form is known as well \cite{MTW3,KMW}.

From relations (\ref{DA}) and (\ref{S2simp}) it is apparent that the
spin two structure function consists of a de Sitter invariant part
plus a de Sitter breaking part which is cubic in $u \equiv \ln(a a')$
\cite{MTW3,KMW},
\begin{equation}
\mathcal{S}_2(x;x') = S_2(y) + \delta S_2(u) \; .
\end{equation}
We shall not require the explicit result of acting the projectors
on the de Sitter invariant part. The de Sitter breaking part gives
\cite{KMW},
\begin{eqnarray}
\lefteqn{i\Bigl[\mbox{}_{\mu\nu} \Delta^{\rm br,2}_{\rho\sigma}
\Bigr](x;x') \!=\! \frac1{4 H^4} \mathbf{P}_{\mu\nu}^{~~\alpha\beta}(x)
\! \times\! \mathbf{P}_{\rho\sigma}^{~~\kappa\lambda}(x') \Bigl[
\mathcal{R}_{\alpha\kappa}(x;x') \mathcal{R}_{\beta\lambda}(x;x')
\delta S_2(u)\Bigr] \; , } \\
& & \hspace{-.5cm} = k \Biggl[ \ln(4 a a') \!+\! 2 \psi\Bigl(
\frac{D \!-\!1}2\Bigr) \!-\! 4 \!+\! \frac1{D \!-\! 1}\Biggr]
(a a')^2 \Biggl\{ 2 \overline{\eta}_{\mu (\rho}
\overline{\eta}_{\sigma) \nu} \!-\! \frac2{D \!-\! 1} \,
\overline{\eta}_{\mu\nu} \overline{\eta}_{\rho\sigma} \Biggr\} ,
\qquad \label{dS2br}
\end{eqnarray}
where $\eta_{\mu\nu}$ is the spacelike Lorentz metric and an overbar
denotes the suppression of its temporal components,
\begin{equation}
\overline{\eta}_{\mu\nu} \equiv \eta_{\mu\nu} + \delta^0_{\mu}
\delta^0_{\nu} \; .
\end{equation}

\section{The Computation}\label{compute}

In this section we assemble the results which have just been
to compute the linearized Weyl-Weyl correlator for the de
Donder gauge propagator. The argument consists of five steps.
One paragraph is devoted to each step.

The first step consists of expressing the linearized Weyl-Weyl
correlator in terms of the graviton propagator. We define the
graviton field by expanding the full metric around de Sitter,
\begin{equation}
\Bigl({\rm full metric}\Bigr)_{\mu\nu}(x) \equiv g_{\mu\nu}(x) +
\sqrt{16 \pi G} \, h_{\mu\nu}(x) \; .
\end{equation}
Because the Weyl tensor of de Sitter vanishes, the Weyl tensor
of the full metric is linear in the graviton field. It can be
given a very simple form using the tensors defined in expressions
(\ref{scriptR}) and (\ref{scriptC}),
\begin{equation}
C_{\alpha\beta\gamma\delta}(x) =
\mathcal{C}_{\alpha\beta\gamma\delta}^{~~~~~ \epsilon\zeta\kappa\lambda}
\times \mathcal{R}_{\epsilon\zeta\kappa\lambda}^{~~~~ \theta\phi\psi\omega}
\times -\frac12 D_{\phi} D_{\omega} \sqrt{16 \pi G} \, h_{\theta\psi}(x) +
O(G) \; .
\end{equation}
The graviton propagator is the expectation value (the time-ordered
product) of two graviton fields. Hence the linearized Weyl-Weyl
correlator can be expressed using the operations of Riemannization
and Weylization that were defined in expressions (\ref{Rdef}) and
(\ref{Weylization}),
\begin{eqnarray}
\lefteqn{\Bigl\langle \Omega \Bigl\vert C_{\alpha\beta\gamma\delta}(x)
\times C_{\mu\nu\rho\sigma}(x') \bigr\vert \Omega \Bigr\rangle }
\nonumber \\
& & \hspace{1cm} = 4\pi G \, {\rm Weyl}\Biggl( {\rm Riem}\Biggl[
D_{\beta} D_{\delta} D'_{\nu} D'_{\sigma} \, i\Bigl[
\mbox{}_{\alpha\gamma} \Delta_{\mu\rho}\Bigr](x;x') \Biggr] \Biggr)
+ O(G^2) \; . \qquad \label{step1}
\end{eqnarray}

The second step is to note that the spin zero part of the propagator
drops out of the Weyl-Weyl correlator. To see this, first write
(\ref{step1}) as,
\begin{eqnarray}
\lefteqn{\Bigl\langle \Omega \Bigl\vert
C_{\alpha\beta\gamma\delta}(x) \times C_{\mu\nu\rho\sigma}(x')
\bigr\vert \Omega \Bigr\rangle } \nonumber \\
& & \hspace{1.4cm} = 4 \pi G \,
\mathcal{P}_{\alpha\beta\gamma\delta}^{~~~~~ \kappa\lambda}(x)
\times \mathcal{P}_{\mu\nu\rho\sigma}^{~~~~~ \theta\phi}(x') \times
i\Bigl[ \mbox{}_{\kappa\lambda} \Delta_{\theta\phi}\Bigr](x;x') +
O(G^2) \; , \qquad
\end{eqnarray}
where $\mathcal{P}_{\alpha\beta\gamma\delta}^{~~~~ \kappa\lambda}$
is the contraction of (\ref{scriptC}) and (\ref{scriptR}) into two
covariant derivatives,
\begin{equation}
\mathcal{P}_{\alpha\beta\gamma\delta}^{~~~~~ \theta\psi} \equiv
\mathcal{C}_{\alpha\beta\gamma\delta}^{~~~~~
\epsilon\zeta\kappa\lambda} \times
\mathcal{R}_{\epsilon\zeta\kappa\lambda}^{~~~~ \theta\phi\psi\omega}
\times D_{\phi} D_{\omega} \; . \label{scriptP}
\end{equation}
Note that the differential operator (\ref{scriptP}) projects out
both longitudinal and trace terms,
\begin{equation}
\mathcal{P}_{\alpha\beta\gamma\delta}^{~~~~~ \kappa\lambda}
D_{\kappa} = 0 = \mathcal{P}_{\alpha\beta\gamma\delta}^{~~~~~
\kappa\lambda} g_{\kappa\lambda} \; .
\end{equation}
Of course this means it annihilates the spin zero projector
(\ref{spin0op}). Hence the linearized Weyl-Weyl correlator derives
entirely from the spin two part of the propagator (\ref{Spin2}),
\begin{eqnarray}
\lefteqn{\Bigl\langle \Omega \Bigl\vert
C_{\alpha\beta\gamma\delta}(x) \times C_{\mu\nu\rho\sigma}(x')
\bigr\vert \Omega \Bigr\rangle } \nonumber \\
& & \hspace{1.4cm} = 4 \pi G \,
\mathcal{P}_{\alpha\beta\gamma\delta}^{~~~~~ \kappa\lambda}(x)
\times \mathcal{P}_{\mu\nu\rho\sigma}^{~~~~~ \theta\phi}(x') \times
i\Bigl[ \mbox{}_{\kappa\lambda} \Delta^2_{\theta\phi}\Bigr](x;x') +
O(G^2) \; . \qquad \label{step2}
\end{eqnarray}

The next step is to note that the de Sitter breaking contribution to
the spin two part drops out. From the conformal invariance of the
Weyl tensor we have,
\begin{eqnarray}
\lefteqn{{\rm Weyl}\Biggl( {\rm Riem}\Biggl[ D_{\beta} D_{\delta}
D'_{\nu} D'_{\sigma} \, i\Bigl[ \mbox{}_{\alpha\gamma}
\Delta^{\rm br,2}_{\mu\rho}\Bigr](x;x') \Biggr] \Biggr) } \nonumber \\
& & \hspace{1cm} = (a 'a)^2 {\rm Weyl}\Biggl( {\rm Riem}\Biggl[
\partial_{\beta} \partial_{\delta} \partial'_{\nu} \partial'_{\sigma}
\, \Bigl\{ (a a')^{-2} \, i\Bigl[ \mbox{}_{\alpha\gamma} \Delta^{\rm
br,2}_{\mu\rho}\Bigr](x;x') \Bigr\} \Biggr] \Biggr) \; . \qquad
\end{eqnarray}
Now use expression (\ref{dS2br}) for the de Sitter breaking
contribution to the spin two part of the propagator to conclude,
\begin{eqnarray}
\lefteqn{ \partial_{\beta} \partial_{\delta} \partial'_{\nu}
\partial'_{\sigma} \, \Biggl\{ (a a')^{-2} \, i\Bigl[
\mbox{}_{\alpha\gamma} \Delta^{\rm br,2}_{\mu\rho}\Bigr](x;x')
\Biggr\} } \nonumber \\
& & \hspace{.5cm} = \Biggl\{ 2 \overline{\eta}_{\alpha (\mu}
\overline{\eta}_{\rho ) \gamma} \!-\! \frac{2
\overline{\eta}_{\alpha\gamma} \overline{\eta}_{\mu\rho}}{D \!-\! 1}
\Biggr\} \times
\partial_{\beta} \partial_{\delta} \partial'_{\nu}
\partial'_{\sigma} \Bigl[ \ln(4 a a') + {\rm Constant}\Bigr] = 0 \;
. \qquad \label{dSbreak}
\end{eqnarray}
Hence the linearized Weyl-Weyl correlator derives entirely from the
de Sitter invariant contribution to the spin two part,
\begin{eqnarray}
\lefteqn{\Bigl\langle \Omega \Bigl\vert
C_{\alpha\beta\gamma\delta}(x) \times C_{\mu\nu\rho\sigma}(x')
\bigr\vert \Omega \Bigr\rangle = 4 \pi G \,
\mathcal{P}_{\alpha\beta\gamma\delta}^{~~~~~ \epsilon\zeta}(x)
\times \mathcal{P}_{\mu\nu\rho\sigma}^{~~~~~ \theta\phi}(x') }
\nonumber \\
& & \hspace{.5cm} \times \frac1{4 H^4}
\mathbf{P}_{\epsilon\zeta}^{~~\kappa\lambda}(x) \! \times\!
\mathbf{P}_{\theta\phi}^{~~\psi\omega}(x') \Bigl[
\mathcal{R}_{\kappa\psi}(x;x') \mathcal{R}_{\lambda\omega}(x;x')
S_2(y)\Bigr] + O(G^2) \; . \qquad \label{step3}
\end{eqnarray}

In step four we take advantage of the fact that
$\mathcal{P}_{\alpha\beta\gamma\delta}^{~~~~~ \epsilon\zeta}(x)$
projects out longitudinal and trace parts to apply identity
(\ref{1stPsimp}) twice in expression (\ref{step3}),
\begin{eqnarray}
\lefteqn{\Bigl\langle \Omega \Bigl\vert
C_{\alpha\beta\gamma\delta}(x) \times C_{\mu\nu\rho\sigma}(x')
\bigr\vert \Omega \Bigr\rangle = \frac{\pi G}{4 H^4} \Bigl( \frac{D
\!-\! 3}{D \!-\! 2}\Bigr)^2 \,
\mathcal{P}_{\alpha\beta\gamma\delta}^{~~~~~ \epsilon\zeta}(x)
\times \mathcal{P}_{\mu\nu\rho\sigma}^{~~~~~ \theta\phi}(x') }
\nonumber \\
& & \hspace{.5cm} \times \Biggl[ \mathcal{R}_{\epsilon\theta}
\mathcal{R}_{\zeta\phi} \square \Bigl[\square \!-\! (D \!-\! 2)
H^2\Bigr] \square' \Bigl[\square' \!-\! (D \!-\! 2) H^2\Bigr]
S_2(y)\Biggr] + O(G^2) \; . \qquad \label{step3.5}
\end{eqnarray}
Now note that $\square' F(y) = \square F(y)$ \cite{OW} and use
expression (\ref{S2simp}) to conclude,
\begin{equation}
\square \Bigl[\square \!-\! (D \!-\! 2) H^2\Bigr] \square'
\Bigl[\square' \!-\! (D \!-\! 2) H^2\Bigr] S_2(y) = 32 H^4
\Bigl(\frac{D \!-\! 2}{D \!-\! 3}\Bigr)^2 A(y) \; ,
\end{equation}
where $A(y)$ is the de Sitter invariant part (\ref{DeltaA}) of the
scalar propagator. Substituting this in (\ref{step3.5}) implies,
\begin{eqnarray}
\lefteqn{\Bigl\langle \Omega \Bigl\vert
C_{\alpha\beta\gamma\delta}(x) \times C_{\mu\nu\rho\sigma}(x')
\bigr\vert \Omega \Bigr\rangle } \nonumber \\
& & \hspace{-.5cm} = 8 \pi G \,
\mathcal{P}_{\alpha\beta\gamma\delta}^{~~~~~ \epsilon\zeta}(x)
\!\times \! \mathcal{P}_{\mu\nu\rho\sigma}^{~~~~~ \theta\phi}(x')
\!\times\! \Bigl[ \mathcal{R}_{\epsilon\theta}(x;x')
\mathcal{R}_{\zeta\phi}(x;x') A(y)\Bigr] + O(G^2) \; . \qquad
\label{step4}
\end{eqnarray}

The final step begins by expressing (\ref{step4}) in terms of
Riemannization and Weylization,
\begin{eqnarray}
\lefteqn{\Bigl\langle \Omega \Bigl\vert
C_{\alpha\beta\gamma\delta}(x) \times C_{\mu\nu\rho\sigma}(x')
\bigr\vert \Omega \Bigr\rangle } \nonumber \\
& & \hspace{-.7cm} = 4 \pi G \, {\rm Weyl}\Biggl( {\rm Riem}\Biggl[
D_{\beta} D_{\delta} D'_{\nu} D'_{\sigma} \Bigl[ \Bigl(
\mathcal{R}_{\alpha\mu} \mathcal{R}_{\gamma\rho} \!+\!
\mathcal{R}_{\alpha\rho} \mathcal{R}_{\gamma\mu}\Bigr) A(y) \Bigr]
\Biggr] \Biggr) \!+\! O(G^2) \; . \qquad
\end{eqnarray}
Note that acting any of the covariant derivatives on the
intermediate tensor factors produces a metric \cite{KW2,KMW},
\begin{equation}
D_{\beta} \mathcal{R}_{\alpha\mu}(x;x') = \frac12 g_{\alpha\beta}(x)
\frac{\partial y(x;x')}{\partial {x'}^{\mu}} \; .
\end{equation}
Because any terms of this form are annihilated by Weylization, we
can move the four covariant derivatives through to act on $A(y)$.
Even acting two mixed derivatives erases the difference between
$A(y)$ and the full scalar propagator,
\begin{equation}
D_{\beta} D'_{\nu} A(y) = D_{\beta} D'_{\nu} i\Delta_A(x;x') \; .
\end{equation}
(This is why the de Sitter breaking contribution (\ref{dSbreak}) 
vanished.) Hence we conclude,
\begin{eqnarray}
\lefteqn{\Bigl\langle \Omega \Bigl\vert
C_{\alpha\beta\gamma\delta}(x) \times C_{\mu\nu\rho\sigma}(x')
\bigr\vert \Omega \Bigr\rangle } \nonumber \\
& & \hspace{-.3cm} = 4 \pi G {\rm Weyl}\Biggl( {\rm Riem}\Biggl[
\Bigl( \mathcal{R}_{\alpha\mu} \mathcal{R}_{\gamma\rho} \!+\!
\mathcal{R}_{\alpha\rho} \mathcal{R}_{\gamma\mu}\Bigr) D_{\beta}
D_{\delta} D'_{\nu} D'_{\sigma} \, i\Delta_A \Biggr] \Biggr) \!+\!
O(G^2) \; . \qquad \label{step5}
\end{eqnarray}
Reshuffling some indices gives the same form (\ref{oldweyl}) that
was derived for the older of the two de Sitter breaking propagators
\cite{TW2,RPW}.

Because our result (\ref{step5}) for the linearized Weyl-Weyl
correlator is the same as for the other de Sitter breaking
propagator, taking the coincidence limit and contracting the
indices must reproduce expression (\ref{Csquared}) as well.
It is worth pointing out that this is the first time the de Donder 
gauge propagator has been used in a loop computation.

\section{Discussion}\label{discuss}

We have computed the linearized Weyl-Weyl correlator for the
recently constructed graviton propagator in de Donder gauge
\cite{MTW3,KMW}. Our result (\ref{step5}) is identical to 
expression (\ref{oldweyl}), which was found using the graviton
propagator constructed with a noninvariant gauge fixing term
\cite{TW2,RPW}. Mathematical physicists obtain the same result
\cite{Kouris}, after correcting some mistakes \cite{Atsuchi}.

We do not accept that the Weyl-Weyl correlator provides a
complete check of the graviton propagator. It is sensitive to 
neither the spin zero part nor to the infrared divergent, de 
Sitter breaking part. However, it does offer a partial check, 
and both de Sitter breaking propagators pass this check. 
Perhaps that fact may ease doubts that have been expressed 
about these propagators accessing a different sector of the 
graviton Hilbert space \cite{HMM}.

Turnabout is fair play, so let us suggest an interesting
test of the exact gauge, de Sitter invariant propagators 
which are claimed to result from singular limits of the
provably false \cite{MTW1} procedure of adding de Sitter 
invariant gauge fixing terms \cite{FH}. This is to act the 
graviton kinetic operator on them and then integrate the 
result onto itself to check that it is a functional projection 
operator. We predict that the alleged propagators will fail 
this test. It is worth noting that the de Sitter breaking, 
de Donder gauge propagator \cite{MTW3} was constructed to 
pass it.

Our result provides support for the suspicion that the new de
Donder gauge propagator may be simple to use, in spite of its
cumbersome tensor form (\ref{gravdecomp}-\ref{spin2op}) and 
complicated structure functions. The same sort of cumbersome 
tensors and complicated structure function appear in the 
Lorentz gauge photon propagator \cite{TW7}. However, all 
known loop computations \cite{PTW} result in the tensors 
contracting to simple forms, and in precisely the right 
differential operators being acted to simplify the structure 
function. We so far have only this one result for the new
propagator, but the same sort of simplifications took place
in (\ref{step3.5}). It will be interesting to see what happens
with other computations.

\vskip 1cm

\centerline{\bf Acknowledgements}

We have profited from conversations with S. P. Miao, who is
responsible for having derived the crucial result (\ref{1stPsimp})
\cite{KMW}. We should also mention A. Higuchi and D. Marolf, whose
persistent skepticism about the de Donder gauge propagator
\cite{MTW4} inspired this study. This work was partially supported
by European Union program Thalis ESF/NSRF 2007-2013, by European
Union Grant FP-7-REGPOT-2008-1-CreteHEPCosmo-228644, by NSF grants
PHY-0855021 and PHY-1205591, and by the Institute for Fundamental
Theory at the University of Florida.

\end{document}